\documentstyle[aps,preprint]{revtex}
\begin{document}
\draft
\title{Composite vertices that lead to soft form factors }
\author{L.\ C.\ Liu}
\address{Theoretical Division, Los Alamos National Laboratory, \\
  Los Alamos, NM 87545 }
\author{Q. Haider}
\address{Physics Department, Fordham University, Bronx, NY 10458 }
\author{J.\ T.\  Londergan}
\address{Department of Physics and Nuclear Theory Center,
Indiana University,\\
  Bloomington, IN 47405 }
\date{February 15, 1995}
\maketitle
\begin{abstract}
The momentum-space cut-off parameter
$\Lambda$ of hadronic vertex functions is studied in
this paper. We use a composite model where we can measure the contributions
of intermediate particle propagations to $\Lambda$. We show that in
many cases a composite vertex function has a much smaller cut-off than
its constituent vertices, particularly when light constituents such as
pions are present in the intermediate state. This suggests that composite
meson-baryon-baryon vertex functions are rather soft, i.e., they have
$\Lambda$ considerably less than 1\ GeV. We discuss the origin of this
softening of form factors as well as the implications of our
findings on the modeling of nuclear reactions.
\end{abstract}

\pacs{PACS numbers: 14.20.-c,Gk; 21.30.+y; 25.80.-e}

\narrowtext
\section{Introduction} \label{sec:I}
Hadronic meson-baryon-baryon (MBB) and three-meson (MMM) vertex functions
or form factors are basic inputs to theories of nuclear reactions. They
play an important role in many-body nuclear physics as they provide models
for making specific off-shell extrapolation of elementary amplitudes to
high momenta. In particular, the momentum-space cut-off scale $\Lambda$
of a vertex affects the contribution of a loop diagram in which the vertex
appears. The larger the $\Lambda$ of a given vertex is, the smaller the
momentum cut-off effects will be, and the greater the magnitude of the
loop integration will be for that vertex. It is, therefore, important to
understand the quantities that determine the range of a form factor.

Although hadronic form factors can in principle be calculated microscopically,
the high complexity of the strong-interaction processes involved in vertex
corrections makes such calculations very difficult. As a result, hadronic
form factors are typically parameterized by an ${\it ad\ hoc}$ function.
This function defines the fall-off of form factors with increasing momentum
transfer and is often assumed to depend on a single cut-off parameter,
$\Lambda$. Clearly, this parametric function is not unique. One can use, for
example, a Gaussian or exponential parameterization of the form
factors\ \cite{EJ}.

By far the most commonly used form factor is the multipole form, of which
one specific type is $[\Lambda^{2}/(\Lambda^{2}-q^{2})]^{n}$, with $q^{2}$
being the square of the four-momentum transfer, and $n=1 (2)$ denoting a
monopole (dipole) form factor\ \cite{N1}. The value of $\Lambda$ is then
determined from fitting some specific reaction data. The pheonomenological
nature of this approach creates difficulties in deriving and interpreting
the cut-off parameter of the vertex function. First, because the form
factor serves as a convenient free input to the theories, inadequacies in
the reaction dynamics may be compensated by adjusting $\Lambda$. Second,
in electromagnetic interactions, introduction of form factors leads to
difficulties in maintaining gauge invariance\ \cite{Fria}.

The difficulty in determining the cut-off parameter
 of a form factor is not simply
an academic question, as the value of $\Lambda$ for a given form factor
can depend strongly on the types of reaction data that are being fitted,
as well as on the theoretical model used to deduce that form factor. We
give two examples which illustrate this dependence. The first involves
deduction of pion-nucleon coupling constants and cut-off momenta
from models which
fit nucleon-nucleon scattering data. The second involves inference of
$\Lambda_{\pi NN}$ from deep inelastic scattering (DIS) of leptons from
nucleons.

Many of the features of phenomenological $NN$ interactions were
demonstrated in dispersion relation calculations of correlated and
uncorrelated two-pion-exchange by the Paris and Stony Brook
groups\ \cite{Paris,Jack,JRVW,Durs}.  These calculations showed the
role of multipion exchange in the $NN$ interaction.  In particular,
the dispersion relation treatment highlighted the importance of intermediate
$\Delta$ formation in certain spin-isospin channels.  These models
motivated the development of meson-exchange potentials
(MEP) of the nucleon-nucleon interaction, notably the Bonn
potential\ \cite{Mach,N2}. Such potentials were able to incorporate many
of the important physical effects of the
dispersive calculations, while retaining an ease of computation
which allowed them to be used in calculations of few-body systems.

The cutoff parameter for the Bonn potential, $\Lambda_{\pi NN}\sim 1.3$
GeV, has greatly influenced subsequent calculations of the offshell
behavior of $NN$ amplitudes. This range was necessary to reproduce the
properties of the deuteron and low-energy $NN$ scattering in the Bonn
potential. However, it is now known that such hard
form factors are not a necessary feature of $NN$ interactions, and that
several physical effects argue for a significantly softer form factor.
For example, the J\"ulich group\ \cite{Juli} have shown that
inclusion of $\pi\pi$ interactions allows one to avoid the need for hard
form factors. Janssen {\it et al.}\ \cite{Jans} further showed that the
inclusion of $\pi\rho$ scattering can reduce the value of
$\Lambda_{\pi NN}$.

Thomas and Holinde\ \cite{Holi} showed that one could obtain a good fit
to $NN$ data with a soft $NN\pi$ form factor, if one simultaneously
introduced an additional short-ranged, isovector tensor force.
Recently, Haidenbauer {\it et al.}\ \cite{HHT} have produced a
model where both one- and two-pion exchanges are included explicitly
with soft $NN\pi$ and $\Delta N\pi$ vertices.
Smaller cut-offs are also advocated by an earlier theoretical
investigation\ \cite{Schu} which demonstrates that the boson nature of
the pion requires a cut-off parameter much smaller than 1 GeV. Small
$\Lambda_{\rho NN}$ and $\Lambda_{\rho N\Delta}$ have also been obtained
by Haider and Liu\ \cite{Haid} and by Deister {\em et al.}\ \cite{Deis}
 when composite models are used for the
$\rho NN$ and $\rho N\Delta$ vertices.

Sullivan\ \cite{Sull} showed that nonperturbative contributions from
coupling to mesons could contribute to DIS even at very high energies.
If one assumes such a mechanism, then antiquark contributions can arise
from a process where a nucleon emits a virtual pion and a DIS occurs from the
antiquark in the meson. The mesonic contribution to this process is very
sensitive to the $\pi NN$ form factor. The nonperturbative contribution
from the antiquarks in the meson cannot exceed the total antiquark
contribution, and analyses of DIS including pionic contributions through
the Sullivan mechanism\ \cite{Fran,Kuma} with a monopole form factor give
$\Lambda_{\pi NN}\sim 0.65$ GeV. However, if in addition to pions one
includes other mesons (including strange mesons\ \cite{Thom1}), the
inferred monopole cut-off parameter is then somewhat larger ($\sim$
800\ MeV) \cite{Thom2}, but is still much less than the
1.3\ GeV of the Bonn potential\ \cite{Mach}.

In order to better understand the $\Lambda$
of a hadronic vertex, one must
go beyond the use of phenomenological parameterizations of the form
factor and study the underlying reaction dynamics. As explicit
considerations of reaction mechanisms give rise necessarily to composite
vertex functions, we carry out in this paper a study of the
property of composite vertices. In particular, we investigate a model
which demonstrates how the cut-off momentum $\Lambda$
of a vertex is determined. Our emphasis is on the general
dependence of the $\Lambda$ for a composite vertex function on the masses of
the constitutent particles and the cut-off momenta of its constituent vertices
(or subvertices). The basic analysis is given in Section\ \ref{sec:II}.
We show that when a light constituent like the pion is present, the
composite vertex has a very soft form factor (i.e., small $\Lambda$) even
if the constitutent MBB or MMM form factors are hard (i.e., having very
large $\Lambda$). We further show that the softening of form factors
has a geometrical origin.  Section\ \ref{sec:III} contains
discussion on the implications of our findings on nuclear physics studies.
The conclusions are given in Section\ \ref{sec:IV}.

\section{Composite vertex function} \label{sec:II}
Let us consider the lowest-order composite MBB vertex shown in
Fig.\ \ref{fig1}, where we denote the initial baryon, the final baryon,
and the meson as $a$, $c$, and $d$, respectively. This vertex exemplifies
a composite vertex because the meson $d$ decays into two mesons (denoted 1
and 2) which subsequently interact with the baryons $a$, $b$, and $c$. In
the rest frame of $c$, the four-momenta of these external particles can be
parameterized as $p_{c}=(w,{\bf 0})$, $p_{a}=(a^{0},-{\bf p})$, and
$p_{d}=(w-a^{0}, {\bf p})$. If we denote the four-momentum variable of
the loop integration as $q$, then the momenta of the internal particles
can be parameterized as
$p_{1}=(q^{0};{\bf q})$, $p_{2}=(w-a^{0}-q^{0}, {\bf p-q})$, and
$p_{b}=(a^{0}+q^{0}, {\bf -p+q})$. After projecting out the angular
momentum dependence, the form factor of the composite MBB vertex
function, $V$, is a scalar and depends on two independent
four-momenta which can be chosen, for example, as $p_{a}+p_{d}$ and
$p_{c}-p_{a}$. From these two momenta, one can form three independent
scalar variables. Hence, in the most general case, we can write, for
example, $V=V(p_{a}^{2},s,t)$ with $s=(p_{a}+p_{d})^{2}\equiv w^{2}$
and $t=(p_{c}-p_{a})^{2}$. For simplicity, we shall consider a less
general case in which the particle $a$ is on its mass shell, i.e.,
$p_{a}^{2}=m_{a}^{2}$ or $a^{0}=E_{a}({\bf p})=\sqrt{{\bf p}^{2} +
m_{a}^{2}}$. Consequently, $V$ depends only on two scalar variables
which can be chosen as $s$ and $t$.

For definiteness, we assume s-wave interactions for
the overall $a+d\rightarrow c$ process and for all subprocesses
$d\rightarrow 1+2$, $ 1+a\rightarrow b$, and $2+b\rightarrow c$.
An actual example of such case is the vertex function for $f^{0}NN$
coupling via intermediate $\pi\pi N^{*}(1535)$. However, in the following,
we shall treat the masses $m_{1}$, $m_{2}$, and $m_{b}$ as variables in
order to explore the general feature of a composite vertex function.
We further assume that
all baryons have spin--$\frac{1}{2}$. The composite MBB vertex function
$V_{da;c}$ is then given by
\begin{eqnarray}
 V_{da;c}(s,t)= \frac{i}{(2\pi)^{4}}C_{I} \sum_{\sigma,\sigma '}
\bar{u}^{\sigma'}(p_{c})\int d^{4}q\ \frac{ g_{d;12}(p_{d}^{2})
F_{d;12}(\Lambda_{0}^{2},t_{12})}{q^{2}-m_{1}^{2}+i\varepsilon}
& &  \nonumber \\
\times\frac{ g_{1a;b}(p_{b}^{2})F_{1a;b}(\Lambda^{2}_{1},q^{2}) }{
(p_{d}-q)^{2}
-m_{2}^{2}+i\varepsilon}g_{2b;c}(w)F_{2b;c}(\Lambda^{2}_{2},p_{2}^{2})
\frac{\gamma\cdot (p_{a}+q)+m_{b}}{(p_{a}+q)^{2}-m_{b}^{2}
+i\varepsilon}\ u^{\sigma}(p_{a})\ & ,
\label{eq:1}
\end{eqnarray}
where $C_{I}$ denotes the isospin coefficient, $u^{\sigma}$ and
$\bar{u}^{\sigma'}$ are Dirac spinors, $t_{12}\equiv (p_{1}-p_{2})^{2}/4$,
and $\gamma$ is the Dirac matrix.
In Eq.\ (\ref{eq:1}),
we have followed the convention adopted in the literature by assuming that
the energy-dependence of each vertex function is entirely contained in its
coupling constant $g$, while the residual form factor $F$ depends only on
the momentum cut-off
and the momentum transfer. The coupling constants of the MBB
vertices are denoted by $g_{1a;b}$ and $g_{2b;c}$. They are dimensionless.
The coupling constant of the MMM vertex is $g_{d;12}$, and it has the
dimension of an inverse momentum. For the sake of brevity, we will,
henceforth, denote the coupling constants $g_{d;12}$, $g_{1a;b}$, $g_{2b;c}$
and the associated form factors by $g_{0}$, $g_{1}$, $g_{2}$ and $F_{0}$,
$F_{1}$, $F_{2}$, respectively. In the literature, the $g_{i} (i=0,1,2)$
are further fixed at the values calculated with $p_{d}^{2}=m_{d}^2$,
$p_{b}^{2}=m_{b}^{2}$, $w^{2}=m_{c}^{2}$, respectively. Needless to say,
this latter choice introduces implicitly a specific off-shell behavior of
the vertices.

It is worth emphasizing that the multipole order $n$ of the subvertex
form factors $F$ should be large enough to ensure the convergence of the
integration in Eq.\ (\ref{eq:1}). For s-wave interactions considered above,
it suffices to use monopole
form factors ($n=1$). When higher partial wave interactions ($\ell > 0$) are
present at the subvertices, there will be
a $q^{\ell}$-dependence at the corresponding subvertex. Furthermore,
propagators of particles having spins higher than 1/2
will also introduce higher powers of momenta. In these latter
cases, higher-order multipole form factors $(n>2)$ need to be used to ensure
the convergence of the integration\ \cite{Haid}. Although this complication
introduces more tedious angular momentum algebra, it will not alter the
qualitative conclusion obtained with the present s-wave study (see
Section IV.)

\subsection{Effects of intermediate particle propagations} \label{sec:IIA}
It is instructive to first examine Eq.\ (\ref{eq:1}) in the limit of
$\Lambda_{i}\rightarrow\infty (i=0,1,2)$, which corresponds to having
contact interactions at all the subvertices in  coordinate space.
In this limit, the form factors $F_{0}=F_{1}=F_{2}= 1 $ and
\begin{eqnarray}
 V_{da;c}(s,t)& =& \frac{i}{(2\pi)^{4}}C_{I}\ g_{0}\ g_{1}\ g_{2}\
\sum_{\sigma,\sigma '}\bar{u}^{\sigma'}(p_{c})\int d^{4}q\
\ \frac{ 1}{q^{2}-m_{1}^{2}+i\varepsilon} \nonumber \\
& \times & \left [ \frac{1}{(p_{d}-q)^{2}-m_{2}^{2}+i\varepsilon} \right ]
\frac{\gamma\cdot (p_{a}+q)+m_{b}}{(p_{a}+q)^{2}-m_{b}^{2}+i\varepsilon}\
u^{\sigma}(p_{a})\ .
\label{eq:2}
\end{eqnarray}
To evaluate Eq.\ (\ref{eq:2}), we introduce two Feynman parameters
($\alpha$ and $\beta$) and rewrite it as
\begin{eqnarray}
  V_{da;c}(s,t)= \frac{i}{(2\pi)^{4}}C_{I}\ g_{0}\ g_{1}\ g_{2}\
\sum_{\sigma,\sigma '}\bar{u}^{\sigma'}(p_{c})\ \Gamma(3)\int_{0}^{1}
d\alpha\int_{0}^{1-\alpha}d\beta\int_{-\infty}^{+\infty}
d^{4}q\ &  \nonumber \\
 \times\ \frac{\gamma\cdot (p_{a}+q)+m_{b}}{\left\{\alpha[q^{2}-m_{1}^{2}]
+\beta[(p_{d}-q)^{2}-m_{2}^{2}]+(1-\alpha-\beta)[(p_{a}+q)^{2}-m_{b}^{2}]
+i\varepsilon\right\}^{3}}u^{\sigma}(p_{a})\ .
\label{eq:3}
\end{eqnarray}
Introducing the new integration variable $q'=q+A$ with
$A\equiv (1-\alpha-\beta)p_{a}-\beta p_{d}$, we obtain
\begin{eqnarray}
 V_{da;c}(s,t) & = & \frac{2iC_{I}}{(2\pi)^{4}}\ g_{0}\ g_{1}\ g_{2}
\sum_{\sigma,\sigma '}\bar{u}^{\sigma'}(p_{c})\int_{0}^{1}
d\alpha\int_{0}^{1-\alpha}d\beta
\nonumber \\
& \times & \int_{-\infty}^{+\infty}d^{4}q'
 \frac{\gamma\cdot (p_{a}+q'-A)+m_{b}}{\left(q'^{2}- D^{2}
+i\varepsilon\right)^{3}}u^{\sigma}(p_{a})\ ,
\label{eq:4}
\end{eqnarray}
where we have defined
\begin{equation}
 D^{2} = A^{2} - [(1-\alpha-\beta)(p_{a}^{2}-m_{b}^{2}) +\beta
p_{d}^{2} -\alpha m_{1}^{2}-\beta m_{2}^{2}] \ .
\label{eq:4a}
\end{equation}
Because of the symmetric integration on $q'$, the linear term
$\gamma\cdot q'$ in Eq.\ (\ref{eq:4}) does not contribute. The
$q'$--integration can then be performed with the aid of the
relation\ \cite{Schw}
\begin{equation}
\int \frac{d^{4}q'}{[q'^{2}-D^{2}+i\varepsilon]^{3}}= -\frac{i{\pi}^2}
{2(D^{2} -i\varepsilon)}\ .
\label{eq:5}
\end{equation}
We have, therefore,
\begin{equation}
 V_{da;c}(s,t)= \frac{\pi^{2}}{(2\pi)^{4}}C_{I}\ g_{0}\ g_{1}\ g_{2}\
 \sum_{\sigma,\sigma '}\bar{u}^{\sigma'}(p_{c})\int_{0}^{1}d\alpha
 \int_{0}^{1-\alpha}d\beta \frac{\gamma\cdot (p_{a}-A)+m_{b}}{ D^{2}
 - i\varepsilon } u^{\sigma}(p_{a})\ .
\label{eq:6}
\end{equation}
For $w < m_{b}+m_{2}$, one has $D^{2} > 0.$ Hence, the $i\varepsilon$ can
be dropped and Im$(V)$=0.

We introduce the function
\begin{equation}
 R(s,p)=\frac{V_{da;c}(s,{\bf p}^{2})}{V_{da;c}(s,{\bf p}^{2}=0)} \ ,
\label{eq:6a}
\end{equation}
and define the momentum-space cut-off, $\Lambda$, of the composite vertex
function as the value of $\mid{\bf p}\mid$ such that
\begin{equation}
R(s,\Lambda ) = \frac{V(s,\Lambda^{2})}{V(s,0)} = \frac{1}{2} \ .
\label{eq:7}
\end{equation}
In Fig.\ \ref{fig2}, we show the $R(s,p)$ calculated at $s=m_{N}^{2}$ and
$m_{a}=m_{N}$(=939 MeV). Three sets of intermediate particle masses
were used. Set I (the solid curve) corresponds to
$m_{1}=m_{2}=m_{\pi}$(=139.6 MeV) and $m_{b}=m_{N}$. Set II
(the dashed curve) is obtained with $m_{1}=m_{2}=m_{\pi}$ but
with $m_{b}=m_{N^{*}}$(=1535 MeV). Finally, Set III (the dot-dashed
curve) is the result of using $m_{1}=m_{2}=2m_{\pi}$ and $m_{b}=m_{N}$.
Figure 2 indicates that even when all the cut-offs associated with the
subvertices
($\Lambda_{0}, \Lambda_{1}, \Lambda_{2}$)
are infinite, the momentum-space cut-off $\Lambda$
of the composite vertex function is finite.
Furthermore, $\Lambda$ increases when the masses of the intermediate
particles increase. This finite $\Lambda$ reflects the nonlocality
of the $d+a\rightarrow c$ interaction in coordinate space, arising
from the propagation of the intermediate particles. The more massive
these particles are, the shorter distances they will travel. Consequently,
the nonlocality will decrease and $\Lambda$
will increase. We have summarized the results in Table\ \ref{table1}.

\subsection{Effects of finite momentum-space cut-offs of constituent
vertices} \label{sec:IIB}
To examine quantitatively the effects of finite momentum-space
cut-offs of subvertices,
one must specify the form factors $F_{i} (i=0,1,2)$.  While it is tempting
to employ covariant form factors in their simplest form, e.g., in a
monopole form $\Lambda_{i}^{2}/(\Lambda_{i}^2-q'^{2})$, a word of caution
for using such simple parameterization is in order. We recall that the
monopole form factor and its multipole variants have their origin in
mechanisms based on $t$-channel ``pole dominance." They have been very
successful in fitting experimental data in the spacelike region where
$-q'^{2} > 0$. However, it is also well known that in the timelike region
where $-q'^{2} < 0$, this simple form no longer holds and dispersion
relations have to be used to write down the relevant form factor\ \cite{Grif}.
Because in a loop integration both these regions can be reached at a
subvertex, the use of one single covariant multipole form factor for a
given subvertex, therefore, becomes problematic. In particular, it can
introduce spurious singularities in $q'^{0}$. To avoid such difficulties,
the ``static approximation'' has often been invoked to drop the dependence
on $q'^{0}$ and to use accordingly the resulting noncovariant form factors
$[\Lambda_{i}^{2}/(\Lambda_{i}^{2}+{\bf q'}^{2})]^{n}$, with $n=1 (2)$
for monopole (dipole)\ \cite{Mach,Jans}. The use of noncovariant form
factors will necessarily make the calculations frame-dependent and break
the crossing symmetry property of the composite vertex function. We have
estimated this covariance breaking by comparing the $V_{da;c}(s,t)$
evaluated separately at the rest frames of the particles $a$ and $c$.
We have found that the effect of covariance breaking is about $15\%$.
In view of the phenomenological nature of the hadronic form factors used
in the literature, we consider the static approximation as acceptable.
We will, therefore, use the following noncovariant parameterizations:
\begin{equation}
F_{0} = \frac{\Lambda^{2}_{0}}{\Lambda^{2}_{0}+({\bf p}/2-{\bf q}')^{2}}\ ,
\label{eq:9}
\end{equation}
\begin{equation}
F_{1} =  \frac{\Lambda^{2}_{1}}{\Lambda^{2}_{1}+{\bf q}'^{2}}\ ,
\label{eq:10}
\end{equation}
and
\begin{equation}
F_{2} = \frac{\Lambda^{2}_{2}}{\Lambda^{2}_{2}+({\bf p}-{\bf q}')^{2}}\ .
\label{eq:11}
\end{equation}
We are thus led to evaluate Eq.\ (\ref{eq:4}) in the following form:
\begin{eqnarray}
 V_{da;c}(s,t) & = & \frac{2iC_{I}}{(2\pi)^{4}}\ g_{0}\ g_{1}\ g_{2}
\sum_{\sigma,\sigma '}\bar{u}^{\sigma'}(p_{c})\int_{0}^{1}d\alpha
\int_{0}^{1-\alpha}d\beta
\nonumber \\
& \times & \int_{-\infty}^{+\infty}d^{4}q' \ F_{0}
F_{1}\ F_{2}\
 \frac{\gamma\cdot (p_{a}+q'-A)+m_{b}}{\left(q'^{2}- D^{2}
+i\varepsilon\right)^{3}}u^{\sigma}(p_{a})\ ,
\label{eq:12}
\end{eqnarray}
where the $F_{i} (i=0,1,2)$ are given by Eqs.\ (\ref{eq:9})--(\ref{eq:11}).
We can first perform the $q'^{0}$ integration
in Eq.\ (\ref{eq:12}) either in upper-half or in lower-half complex plane
of $q'^{0}$. Equation\ (\ref{eq:12})
then becomes
\begin{eqnarray}
 V_{da;c}(s,t)= \frac{12}{(2\pi)^{3}} \sum_{\sigma,\sigma '}
\bar{u}^{\sigma'}(p_{c})\int_{0}^{1}d\alpha\int_{0}^{1-\alpha}
d\beta\int d{\bf q'} \ \frac{\gamma\cdot (p_{a}+\bar{q}'-A) +
m_{b}}{(2E({\bf q'}))^{5}} u^{\sigma}(p_{a}) & &  \nonumber \\
\times g_{0}\ g_{1}\ g_{2}\left(\frac{\Lambda^{2}_{0}}{\Lambda^{2}_{0}+
({\bf p}/2-{\bf q'})^{2}}\right)  \left( \frac{\Lambda^{2}_{1}}
{\Lambda^{2}_{1}+{\bf q'}^{2}}\right) \left(\frac{\Lambda^{2}_{2}}
{\Lambda^{2}_{2}+({\bf p}-{\bf q}')^{2}}\right)\ ,
\label{eq:13}
\end{eqnarray}
where $\bar{q}'\equiv (0,{\bf q'})$.

We have evaluated Eq.\ (\ref{eq:13}) in the kinematic region
$s=w^{2}=m_{N}^{2}$ and $ m_{2}+m_{b}> w$ with $m_{1}=m_{2}=m_{\pi}$
and $m_{b}=m_{N}$. Three sets of subvertex cut-offs were used. They are:
$\Lambda_{0}=\Lambda_{1}=\Lambda_{2}=1.2$ GeV (denoted Set IV);
$\Lambda_{0}=0.46$ GeV, $\Lambda_{1}=\Lambda_{2}=1.2$ GeV
(denoted Set V); and  $\Lambda_{0}=0.46$ GeV,
$\Lambda_{1}=\Lambda_{2}=0.65$ GeV (denoted Set VI).
(See also Table\ \ref{table2}.)
The $R(s,p)$ calculated with these three sets are
shown in Fig.\ \ref{fig3} as the
solid, dashed, and dot-dashed curves, respectively.
An inspection of Fig.\ \ref{fig3}
indicates clearly the trend that the $\Lambda$ of the composite vertex
function decreases with the cut-offs of the subvertices. For the purpose
of further examining the effect of intermediate particle masses in the
presence of finite subvertex cut-offs, we also calculated $R(s,p)$ by
keeping $\Lambda_{i} (i=0,1,2)$ to be the same as those of Set IV but
increasing $m_{b}$ from $m_{N}$ to $m_{N^{*}}$. The result is given in
Table\ \ref{table2} as Set VII.

Upon comparing the $\Lambda$ of Set I (respectively, Set II) of
Table\ \ref{table1} with that of Set IV (respectively, Set VII) of
Table\ \ref{table2}, we find that the presence of finite
$\Lambda_{i} (i=0,1,2)$ reduces significantly the $\Lambda$ of the composite
vertex. On the other hand, comparing the $\Lambda$ of Sets IV and VII of
Table\ \ref{table2} shows that it increases with increasing intermediate
particle masses as in the case with infinite $\Lambda_{i}$.

\subsection{Domain of softening of form factors} \label{sec:IIC}
In many cases, the composite vertex is a functional of itself. For
example, one of the three subvertices of the composite $\pi NN$ vertex
considered in Ref.\ \cite{Jans} is the $\pi NN$ vertex itself. The result
$\Lambda < \Lambda_{i}$ for $ i\in (0,1,2)$ then corresponds to a softening
of the form factor of the $i\,$th vertex. We have explored in detail the
dependence of $\Lambda$ on the masses ($m_{1}$, $m_{2}$, $m_{b}$) and the
subvertex cut-offs
 ($\Lambda_{0}$, $\Lambda_{1}$, $\Lambda_{2}$). Here, we present the
results obtained in two of the cases studied. In the first case, we fixed
$m_{b}$ at $m_{N}$ and considered two variables $m^{*}\equiv m_{1}=m_{2}$
and $x\equiv \Lambda_{0}=\Lambda_{1}=\Lambda_{2}$. We calculated the
$\Lambda$ by varying $x$\ and using  three different $m^{*}$. We chose
$m^{*}=$ 140, 550, 770 MeV in order to simulate the masses of light ($\pi$),
medium ($K,\eta$), and heavy ($\rho,\omega$) mesons commonly encountered in
intermediate energy physics. In the second case,  we further fixed $m_{1}$
at $m_{\pi}$. (Fixing $m_{2}$ while varying $m_{1}$ gave essentially the
same results.) In Figs.\ \ref{fig4} and \ref{fig5}, we present the dependence
of
 $\Lambda$ on
$x$ for these two cases. In these two figures, the region under the diagonal
line can be termed the ``domain of softening'' because in this region the
momentum-space cut-off
of the composite vertex is smaller than the cut-offs of the constituent
vertices, i.e., $\Lambda < x$. The region above the diagonal line corresponds
to the reversed situation where $\Lambda > x$. From both figures, we note
that the minimal $x$ (denoted $x_{0}$) for which $\Lambda \leq x$ can
occur increases with meson masses. However, $x_{0}$ is dramatically
reduced when a pion is present in the intermediate state.

We recall that in the composite model of the $\pi NN$ vertex of
Ref.\ \cite{Jans}, the incoming pion decays virtually into a pion
and a rho-meson which subsequently interact with the nucleon. Hence, in
our notation, $m_{1}=m_{\rho}=770$ MeV, $m_{2}=m_{\pi}=140$ MeV, and
$m_{b}=m_{N}$. Further, the cut-off parameters of the constituent vertices in
Ref.\ \cite{Jans} correspond to an $x=1.3$ GeV. From extrapolating the
dot-dashed curve in Fig.\ \ref{fig5}, we see that the parameters used in
that work are situated well within the region where a softening of the
form factor can take place. A similar analysis applies to the appearance
of soft composite $\rho$-baryon-baryon vertices in Ref.\ \cite{Haid}.

\section{Discussion} \label{sec:III}
We have shown the reason and the conditions under which a composite
vertex function can have a momentum-space cut-off  much smaller than its
constit
uent
vertices. We emphasize that our finding is very general. This is because
we have demonstrated that the softening process is closely related to
nonlocalities associated with the ``elementary'' interactions and
the particle propagations. In view of this, we assert that softening of
the form factor will take place in the presence of different types of
meson-baryon interaction so long as the kinematics are favorable.
This assertion is clearly supported by various published
results\ \cite{Jans,Haid}.

When the $i$th subvertex is the composite vertex itself, such as in the
model of the $\pi NN$ vertex considered by Janssen {\it et al.}\ \cite{Jans},
the result $\Lambda < \Lambda_{i}$ can be termed the ``softening'' of the
form factor of the $i$th vertex.  However, under this circumstance, it will
be necessary to iterate the calculations so as to obtain a convergent
solution of $\Lambda$. Such self-consistent calculations are indeed
possible, as has been demonstrated in the study of composite $\rho NN$ and
$\rho N\Delta$ vertices by
Deister {\em et al.}\ \cite{Deis} and by Haider and Liu\ \cite{Haid}, where a
set of coupled integral equations of the composite vertex functions
is established and iterated to all orders.

The Feynman diagram given in Fig.\ \ref{fig1} represents the lowest order
dynamics of a composite vertex function. Clearly, there are many
higher-order diagrams, such as meson loop corrections to the composite
vertex itself\ \cite{Haid} and rescattering between mesons\
\cite{Jans}.
Because higher-order diagrams contain more subprocesses and particle
propagations, each of which increases the overall
coordinate-space nonlocality of the composite vertex, we can anticipate
that inclusion of higher-order diagrams will further decrease the
momentum space cut-off $\Lambda$
of the composite vertex. Indeed, it has been noted in
Ref.\ \cite{Jans} that the inclusion of  $\pi\rho$ rescattering
softens further the composite $\pi NN$ form factor.

It is of interest to examine the analyticity of the composite vertex
function $V(s,t)$. Once the Feynman-diagram representation of the
dynamics is given, calculation of the diagrams will determine both
the real and imaginary parts of the composite vertex function, Re($V$) and
Im($V$). If the Feynman propagators are not approximated, then Re($V$) and
Im($V$) will satisfy causality relations. This is the approach used in this
study. Furthermore, the analytic structure of the propagators ensures
that Im$(V)=0$ in the kinematical regions where the $t$-channel
$d\rightarrow 1+2$ process and the $s$-channel $2+b\rightarrow c$ process
are energetically forbidden.
 An alternative method of relating Re($V$) and Im($V$) is to use
dispersion relations. This is the approach used, for example, in
Refs.\ \cite{Jans} and \cite{Durs},
 where Im($V$) is first calculated from the underlying
reaction diagrams.  The {\it calculated} Im($V$) is then inserted into a
dispersion relation to generate the Re($V$).

It is useful to recall that the advantage of using dispersion relation
resides in the fact that the imaginary part of the scattering amplitude
can be directly obtained from experimental cross sections through the
optical theorem. In other words, the imaginary part is fixed by
experiment and only the real part of the amplitude depends on the
parameters of the theory.  In this regard, we would like to emphasize
that the possibility of using experimental input to obtain the imaginary
part of the composite vertex, Im($V$), is severely limited. This is
because, in general, the overall $da\rightarrow c$ and its
subprocesses are experimentally inaccessible. This experimental
difficulty should not be underestimated, as it will increase considerably
the model dependence of dispersion relation calculations.

We have mentioned that the momentum-space cut-off of a vertex function plays an
important role in the modeling of nuclear reactions as it affects the
magnitude of the contribution by an elementary process to loop diagrams.
In most of the published MEP studies, simple multipole MBB form factors
were used to fit the nucleon-nucleon phase shifts. These fits gave
$\Lambda_{\pi NN}$, $\Lambda_{\pi N\Delta}, \Lambda_{\rho NN},$ and
$\Lambda_{\rho N\Delta}$ of about 1.3 GeV. However, as mentioned in
Section\ \ref{sec:I}, such large $\Lambda_{\pi NN}$ and
$\Lambda_{\pi N\Delta}$ are in conflict with the DIS
results\ \cite{Fran,Kuma,Thom1,Thom2}. In
addition to these difficulties, adverse effects due to the use of large
$\Lambda_{\rho NN}$ and $\Lambda_{\rho N\Delta}$ were also reported in the
analyses of pion production experiments\ \cite{Jain}. While smaller $\Lambda$
have been obtained by means of using composite $\pi NN$, $\rho NN$, and
$\rho N\Delta$ vertex functions\ \cite{Jans,Haid}, it remains an open
question as to the completeness of the physics described by the original
MEP models. One can reasonably expect that when composite, instead of
phenomenological multipole, MBB form factors are used in MEP models to
refit the phase shifts, one or both of the following scenarios could occur.
Either the fit would give increased MBB coupling constants, or more
meson-baryon interaction diagrams would have to be added in order to
compensate numerically the loss of reaction strength, arising from smaller
cut-offs of the composite vertices. In
either of the above scenarios, our understanding of the meson-exchange
dynamics will be significantly improved.
In fact, it has already been noted that
using a soft $\pi NN$ form factor to
fit the $NN$ scattering data requires at least the addition of some more
new diagrams, such as the correlated $\pi\pi$ and
$\pi\rho$ diagrams\ \cite{Juli,Holi,Jans1}.

The use of smaller $\Lambda$ associated with composite hadronic vertices
will equally challenge current fits to other reactions. This is because
most of these best fits were based on the use of
$\Lambda_{MBB}\stackrel{>}{_{\sim}} 1.2 $\ GeV.  We would like to mention,
in particular, the pion-nucleus double charge exchange reactions.
These reactions are still far from being fully understood, and indeed,
high sensitivities of these reactions to the $\pi NN$ cut-offs have
been noted\ \cite{HaL1,HaL2}. In the light of the present study,
we believe that this sensitivity might be used to assess objectively
the importance of certain reaction mechanisms and, in so doing, to
discover new directions for improving the theories.

\section{Conclusions} \label{sec:IV}
Mesonic corrections to elementary processes give rise naturally to
composite vertex functions. Use of these composite vertices is,
therefore, required by microscopic nuclear reaction theories. In most
cases, especially when one of the intermediate particles is a light
meson such as the pion, the momentum-space cut-off of the composite vertex
is much smaller than those of the constituent vertices. When the composite
vertex is a functional of itself, this reduction of the
momentum-space cut-off can be termed
as the softening of the vertex or form factor. However, as has been shown
in Refs.\ \cite{Haid} and \cite{Deis},
it is necessary to carry out self-consistent
calculations in order to ensure the convergence of the iterative calculations.
Clearly, the notion of a composite vertex can be extended to the regime of
very high momentum transfers to include the exchanges of quarks and
gluons among the hadrons.

The softening of form factors has been observed in theoretical calculations
of $\rho$-baryon-baryon\ \cite{Haid} and $\pi$-baryon-baryon
vertices\ \cite{Jans}. Although these vertex calculations involve very
different types of intermediate particles and meson-baryon interactions,
the results obtained all follow the pattern outlined in Section\ \ref{sec:II}.
In particular, one notes from Ref.\ \cite{Haid} that the softening of
composite $\rho$-baryon-baryon vertex functions occur in the presence of
p- and d-wave interactions at various subvertices. This general agreement
with the result given by the present s-wave model strongly supports
our findings on the principal role played by the nonlocality in the
softening of form factors.

The use of composite vertices and the associated small momentum-space
cut-offs will
inevitably lead to a reexamination of many existing fits to various
reactions, which were obtained with form factors having a monopole cut-off
of 1 GeV or greater. On the other hand, this reexamination will aid us to
clarify the underlying physics governing meson-nucleus interaction.

The use of composite vertex functions not only provides a dynamical
realization of soft MBB
form factors ($\Lambda < 1 $GeV), but also naturally extends it
to the domain of complex variables. Indeed, as can be seen from the
analytic structure of the Feynman propagators in Eq.\ (\ref{eq:1}),
for values of $s$ or $t$ above the corresponding inelastic thresholds,
$V(s,t)$ is complex-valued. As pointed out in Ref.\ \cite{Haid}, while
it may be a good approximation to employ a real-valued vertex function in
analyzing nucleon-nucleon scattering below the pion production threshold,
the situation is very different in meson production experiments where
$V$ can be complex-valued. The non-vanishing imaginary part of the
composite vertex functions can introduce interference effects in nuclear
reaction calculations. This new aspect of the nuclear dynamics has
so far not been studied in the literature and definitely merits a
systematic investigation in the future.

\begin{center}
{\bf ACKNOWLEDGEMENTS}
\end{center}

Two of us (Q.H. and L.C.L.) would like to thank the staff of the Nuclear
Theory Center of Indiana University, Bloomington, IN, where some of the
work reported here was carried out, for their hospitality. This work was
done under the auspices of the U.S. Department of Energy and the National
Science Foundation.

\begin{table}
\caption{The cut-offs $\Lambda$ (in GeV) of the composite vertex function
calculated in the kinematic domain
$s=w^{2}=m_{N}^{2}$ and $ m_{2}+m_{b} > w.$
\label{table1}}
\begin{tabular}{cccccccc}
Set & $m_{b}$ & $m_{1}$ & $m_{2}$  & $\Lambda_{0}$ &$\Lambda_{1}$&
$ \Lambda_{2}$ &  $ \Lambda $  \\
\tableline
 I &$ m_{N}$ & $m_{\pi}$ & $ m_{\pi} $ &$\infty$&$ \infty$&$ \infty$& 0.60
 \\
 II &$ m_{N*}$ & $m_{\pi}$ & $ m_{\pi} $&$ \infty$&$ \infty$&$ \infty $& 1.14
  \\
 III &$ m_{N}$ & $ 2m_{\pi}$ & $ 2m_{\pi} $ &$ \infty$&$ \infty$&$ \infty $ &
1.30   \\
\end{tabular}
\end{table}

\begin{table}
\caption{The cut-offs $\Lambda$ (in GeV) of the composite  vertex function
calculated in the presence of finite subvertex cut-offs.
\label{table2}}
\begin{tabular}{cccccccc}
Set & $m_{b}$ & $m_{1}$ & $m_{2}$  & $\Lambda_{0}$ &$\Lambda_{1}$&
$ \Lambda_{2}$ &  $ \Lambda $  \\
\tableline
IV &$ m_{N}$ & $m_{\pi}$ & $ m_{\pi} $ & $ 1.20$&$ 1.20$&$ 1.20$ &0.42
\\
V &$ m_{N}$ & $ m_{\pi}$ & $ m_{\pi} $& $ 0.46$&$ 1.20$&$ 1.20$& 0.36 \\
VI &$ m_{N}$ & $ m_{\pi}$ & $ m_{\pi} $& $ 0.46$&$ 0.65$&$ 0.65$ & 0.31
 \\
VII &$ m_{N*}$ & $m_{\pi}$& $m_{\pi}$ & 1.20 & 1.20 & 1.20& 0.55 \\
\end{tabular}
\end{table}

\begin{figure}
\caption{ A leading-order Feynman diagram for a composite
$d+a\rightarrow c$ vertex. The solid and dashed
lines denote, respectively, the baryons and mesons. The circle
symbolizes the corresponding
phenomenological vertex  which approximates the internal dynamics
by means of a multipole form factor.
\label{fig1}}
\end{figure}

\begin{figure}
\caption{ $R(s,p)$  calculated at
$s=  m_{N}^{2}$ and $m_{a}= m_{N}$ with the parameters given by
Set I (solid curve), Set II (dashed curve), and Set III (dot-dashed
curve) in Table\ I.
\label{fig2}}
\end{figure}

\begin{figure}
\caption{ $R(s,p)$  calculated at
$s=  m_{N}^{2}$ and $m_{a}= m_{N}$ with the parameters given by
Set IV (solid curve), Set V (dashed curve), and Set VI (dot-dashed
curve) in Table\ II.
\label{fig3}}
\end{figure}

\begin{figure}
\caption{ Relation between $\Lambda$ and $x$ at $s=m_{N}^{2}$
and $ m_{a}=m_{b}=m_{N}$.
The solid, dashed, and
dot-dashed curves represent, respectively, the results obtained
with $m^{*}=$ 140, 550, 770 MeV.
\label{fig4}}
\end{figure}

\begin{figure}
\caption{ Relation between $\Lambda$ and $x$ at $s=m_{N}^{2}$,
$m_{a}= m_{b}= m_{N},$ and $m_{1}= m_{\pi}$. The solid, dashed, and
dot-dashed curves represent, respectively, the results obtained
with $m_{2}=$ 140, 550, 770 MeV.  \label{fig5}}
\end{figure}


\begin{references}
\bibitem{N1} It is important to note that slightly different versions of
multipole form factors can correspond to significantly different ``cut-offs.''
In comparing different parameterizations, one must be careful to compare
similar forms, or to note differences from the ``standard'' form we present
here. We will try to note such differences in this paper.
\bibitem{N2} Note that the Bonn group uses the non-covariant form factor
$[(\Lambda^{2}-m^{2})/(\Lambda^{2}-{\bf q}^{2})]^{n}$ where $m$ is the mass
of the exchanged meson and ${\bf q}$ is the three-momentum transfer. One
must exercise care in comparing the cut-off of this form factor with ours.
\end{references}
\end{document}